\documentclass[twocolumn,prl,aps,groupedaddress]{revtex4}
\usepackage{graphicx}
\begin{document}

\title{Pouring Excitons into a Two-Dimensional Bucket: Equilibration of Indirect Excitons in an In-Plane Harmonic Potential}
\author{Z. V\"or\"os,$^1$ D.W. Snoke,$^1$ L. Pfeiffer,$^2$ and K. West$^2$}
\affiliation{$^1$Department of Physics and Astronomy, University of Pittsburgh, Pittsburgh, PA 15260\\
$^2$Bell Labs, 700 Mountain Ave., Lucent Technologies, Murray Hill, NJ  07974}

\begin{abstract} We have trapped a gas of long-lifetime, high-mobility excitons in an in-plane
harmonic potential. Trapping is an important step toward the goal of a controlled Bose-Einstein condensate of excitons. We show that the repulsive interaction between the excitons plays a dominant role in the behavior of the excitons, in contrast to the weak interactions in atomic gases. We show that under proper conditions the excitons thermalize in the trap to a well-defined equilibrium spatial distribution.
\end{abstract}

\maketitle

\vspace{.5cm}

\begin{figure}[b]
\includegraphics[width=0.45\textwidth]{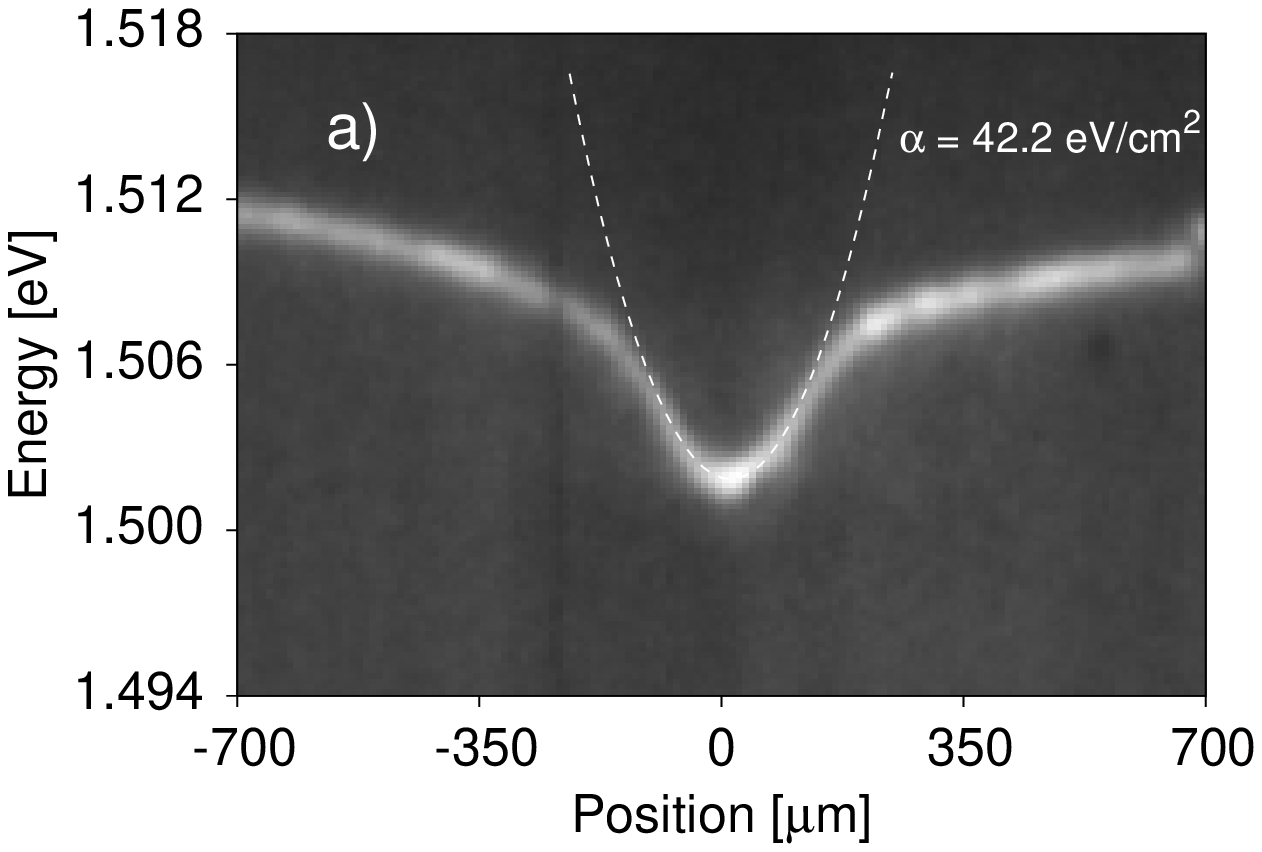}
\includegraphics[width=0.45\textwidth]{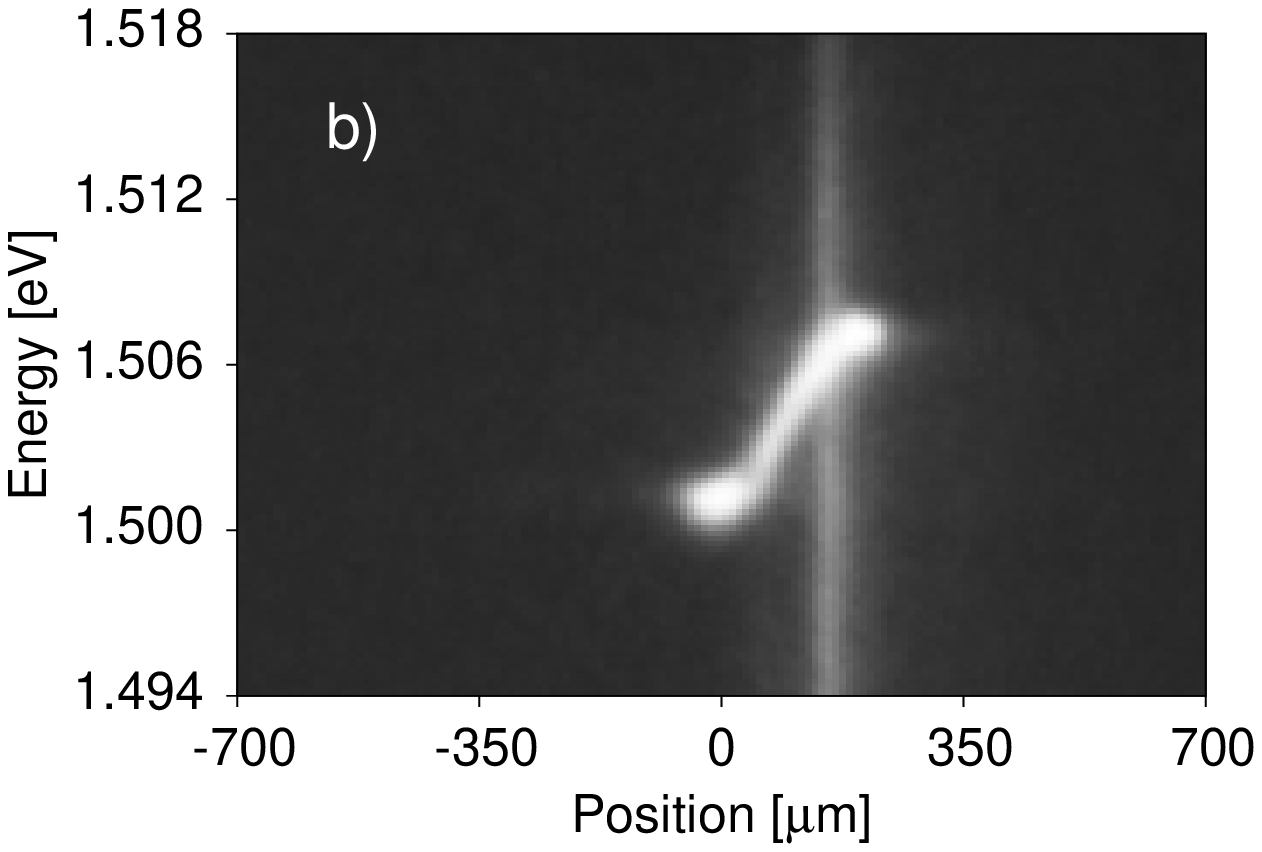}
\caption{The exciton ``bucket.'' a) The potential profile, obtained by 
illuminating the entire area with a defocused, very low density laser. The light emitted by the excitons gives the energy of the excitons at each point. Dashed line: best fit of the function $U = \frac{1}{2}\alpha r^2$ to the intensity maximum of the luminescence at each point. b)  Excitons flowing into the bucket, created by a tightly focused laser at $r=200$ $ \mu$m, with all other conditions the same. The vertical line is substrate luminescence at the point of the laser focus (since the substrate is heavily doped, a series of impurity luminescence lines appear below the band gap).  Indirect excitons flow down from this point of creation to the bottom of the well.  }
\end{figure}

Although Bose-Einstein condensation (BEC) of excitons has been theoretically predicted for over 40 years \cite{science-review}, there is no experimental example of a controlled, well-defined exciton condensate in a trap analogous to atom BEC in a trap. In many exciton condensate experiments (and the related system of exciton-polaritons), excitons have been created and observed in a highly nonequilibrium state by a laser pulse, in some cases with coherence of the laser coupling directly to the excitons, and in some cases with unknown spatial boundary conditions, freely diffusing with no defined equilibrium volume, or localized in a disordered landscape. Such conditions make it nearly impossible to theoretically analyze the behavior of  the excitons.

To follow a path analogous to the atom BEC experiments, several experimental milestones must be established. First, one must have a system with exciton lifetime nearly infinite compared to the equilibration time, so that an equilibrium description applies. Second, it must be demonstrated that these excitons move freely as a gas and are not localized by disorder. Third, a method of trapping the excitons in a macroscopic potential minimum must be demonstrated. Fourth, it must be demonstrated that the excitons can equilibrate in this trap, in other words, that they move through the entire potential and equilibrate both spectrally and spatially within their lifetime. Last, the temperature must be low enough, or the density high enough, for actual condensation to occur. In this case, several fascinating effects are theoretically predicted to occur, namely coherent light emission without lasing \cite{coh}, spectral and spatial narrowing of the light emission \cite{levitov}, spontaneous spin polarization \cite{tejedor}, and superfluid transport response \cite{yudson,littlewood-mag}. 

A method of achieving the first experimental goal has been known for  the past decade, which is to create indirect excitons in coupled quantum wells \cite{indirect}. An electric field normal to the planes pulls the electrons and holes into adjacent wells, where they are close enough to feel the Coulomb attraction and form excitons but have greatly reduced decay rate since they must tunnel through barrier to recombine \cite{szymanska}.   In the present experiments, the lifetime of the excitons can be tuned up to 50 $\mu$s. 

The second experimental goal has been accomplished recently with high quality structures, in which diffusion lengths of hundreds of microns have been demonstrated  \cite{diffprl,gart}. The third goal, a method of trapping, was already demonstrated in 1999 \cite{apl}, but at that time the diffusion length of the excitons was too small for the excitons to equilibrate in the trap within their lifetime. We now report successful demonstration of the fourth goal, namely, equilibration of high-mobility excitons in a trap. This is a major step toward an equilibrium Bose-Einstein condensate. To prove that the excitons reach equilibrium, we have performed a comprehensive series of spatially-resolved, spectrally-resolved, and time-resolved measurements using high-mobility samples and improved trapping methods \cite{diffprl}.  Recent work \cite{eisenstein,ibm} on permanent excitons in a two-dimensional electron gas in coupled quantum
wells is a promising complementary approach, resembling a BCS paired state, but this system cannot
undergo true Bose-Einstein condensation since this is forbidden in two dimensions unless there is a
confining potential \cite{bectheory}. Trapping excitons in a harmonic potential also means that we have
a way of studying the interactions of excitons under controlled circumstances, since the
exciton-exciton interactions determine the spectral energy shifts of the luminescence from the
excitons. 

Our double quantum well samples, grown by means of molecular beam
epitaxy (MBE) on $n$-doped GaAs (100) substrates with $p$-type capping layer, consist of two 100-\AA~
GaAs wells separated by a 40-\AA~ $\mathrm{Al}_{0.3}\mathrm{Ga}_{0.7}$As barrier, with the same
properties as discussed in Ref. \cite{diffprl}.  The indirect exciton binding energy is calculated for this structure to be 3.5 meV in the high-field limit \cite{szymanska}. Voltage was applied perpendicular to the wells by contacts to the heavily doped substrate and capping layers. 

Electrons were excited into the conduction band by a laser pulse tuned in wavelength to the direct, single-quantum-well absorption line, to reduce the amount of excess heat given to the lattice as much as possible. The repetition period of the laser was 4 or 8 $\mu$s, so that few excitons remain from previous pulses. After the pulse, under the influence of the
electric field, electrons and holes tunnel into different wells,  producing spatially indirect excitons.  The
transport of indirect excitons was measured by imaging the light from the sample onto the entrance slit of a spectrometer with time-resolved single-photon counting.  All
measurements were conducted in a Janis Varitemp cryostat, with the samples either immersed in liquid He or in helium vapor. 

Fig.~1 shows images of the time-integrated photoluminescence (PL) emitted by the excitons, recorded with an imaging spectrometer with a Princeton intensified CCD camera, giving the photon
energy as a function of position on the sample, for two different cases. The voltage across the wells
was high enough that the indirect exciton energy shifted to below the GaAs band gap. A trap was created using the method of Ref. \cite{apl}, using hydrostatic expansion to create a band-gap minimum. To obtain deeper and narrower traps, the substrate was thinned to 100 $\mu$m, and a pin with approximately 50 $\mu$m tip diameter was pressed against the back side of the substrate to create the trap.   In Fig.~1(a), the laser was defocused, illuminating the entire area. From this image, we can determine the effective spring constant using a parabolic fit to the potential minimum, which as seen in this figure, is a good approximation of the potential over a region of around 300 $\mu$m. In Fig.~1(b), the laser is tightly focused, far from the center of the trap.  As seen in this figure, the excitons flow over 200 $\mu$m from the point of creation down to the energy minimum. 

\begin{figure}
\includegraphics[width=0.45\textwidth]{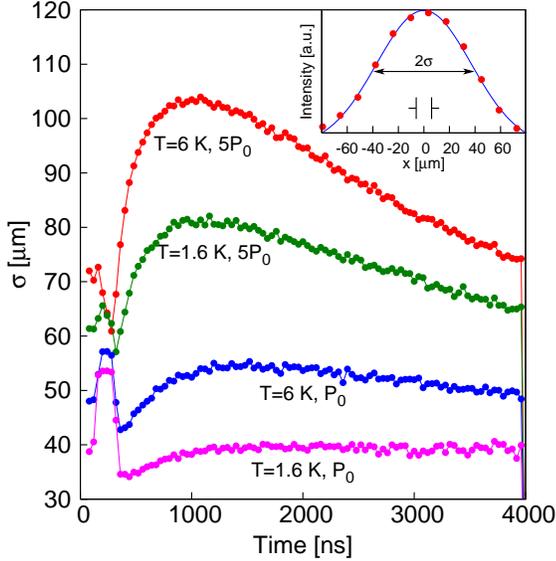}
\caption{Inset: Spatial profile of the exciton luminescence, integrated spectrally,  from 3.9 to 4.0 $\mu$s
after the laser pulse ($P_0 = 0.1$ nJ) resonant with the direct, single-quantum-well exciton absorption line. The solid line
is the best fit to a Gaussian distribution. Lower curves:  The variance of the cloud size as a function of time, taken from the best fit to a Gaussian spatial profile at all times, for two temperatures and two laser pulse energies, as indicated. The data before 0.3 $\mu$s are obscured by bright substrate luminescence, during the 100-ns laser pulse. }
\end{figure}

The inset of Fig.~2 shows a typical spatial profile a long time after the laser pulse, for very low exciton density, when the excitons are created by a tightly focused laser spot in the center of the trap. The main curves of Fig.~2 show the cloud size as a function of time, for two excitation densities and two temperatures.  As seen in this figure, the cloud size at high density initially expands quickly, due to the strong dipole-dipole repulstion of the indirect excitons. At late times, the cloud contracts, since the density drops due to exciton recombination, and therefore the outward pressure decreases. 

In equilibrium, one expects the cloud profile to be given by
\begin{equation}
n(r) = n_0 \exp\left[-\frac{\left(\frac{1}{2}\alpha r^2 +V[n(r)] ]\right) }{k_BT} \right],
\label{equil}
\end{equation}
where $\frac{1}{2}\alpha r^2$ is the harmonic approximation of the externally applied potential. The value of $\alpha$ in
this case was 42 eV/cm$^2$, taken from the best fit to the image data shown in Fig.~1(a). 
$V[n(r)]$ is the energy due to the exciton-exciton interaction, which is proportional to the local exciton density, as discussed below. When the density is low, this term will
drop out, and the spatial profile will be a simple Gaussian, with width given by the temperature and the force constant
of the harmonic potential. The width of the cloud can therefore be plotted in terms of the effective temperature,
$
T_{\rm eff} = \alpha\sigma^2/k_B,
$
where $\sigma$ is the variance of the Gaussian profile.

\begin{figure}
\includegraphics[width=0.45\textwidth]{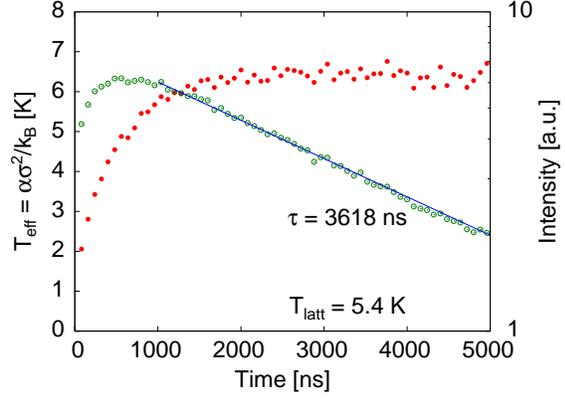}
\caption{Solid circles: the temperature deduced from the cloud size, using the formula given in the text, as a function of time, for lattice temperature $T=5.4$ K and very low laser pulse energy ($P_0 = 0.1$ nJ), which we estimate corresponds to an exciton density of $10^9$ cm$^{-2}$ at the latest times after the laser pulse. Open circles: the total spectrally integrated PL intensity for the indirect excitons as a function of time, for the same data set.}
\end{figure}

In Fig.~3, we show this effective temperature as a function of time in the case of very low density. At long times after the laser pulse,  the cloud size stays constant in time, even as
the exciton density drops, indicating that density-dependent effects are negligible, and that the cloud has reached
thermal equilibrium. Thermal equilibrium is also indicated by the lifetime data.  The rate of photon emission from excitons in GaAs is known \cite{Trate} strongly temperature dependent, and inversely proportional to their temperature in the high-temperature limit, since only excitons with momentum near zero can emit a photon and conserve energy and momentum. If the temperature is changing, the excitons will exhibit non-exponential decay. Fig.~3 shows that at all times after 1 $\mu$s, the total PL intensity of the indirect excitons is single exponential, indicating that the temperature is not changing, consistent with the constant spatial volume at the same times.

Fig.~4 shows that the exciton temperature deduced from the spatial profile agrees well with the helium bath temperature for temperatures above 5 K. At lower temperatures, however, the exciton temperature deviates from the bath temperature, never dropping below 5 K. 

This might not seem surprising at first, because carrier temperatures different from the lattice temperature are common in semiconductor physics,  due to the decreasing rate of phonon emission as the carriers cool. In the present experiments, however, the lifetime of the excitons is so long that they should be well thermalized to the lattice. The time scale for thermalization of the exciton gas to the lattice by phonon emission is well known for excitons
in quantum wells; recent calculations \cite{ivanov}, which have been reproduced for our experimental conditions using
the methods described in earlier work \cite{snokebc}, indicate that the time constant for cooling to the lattice temperature is around 50 ns, far less than the lifetime of the excitons here. The lattice temperature should also be very close to the bath temperature. The total energy deposited by the laser pulse is low, and the time constant for heat diffusion so fast, that the increase of lattice temperature due to excess energy of the laser pulse should be negligible at late times. Because we used a 100 ns, semi-continuous laser pulse, two-photon absorption is negligible. One possibility is that the DC tunneling current through the heterostructure, approximately 1 $\mu$A/cm$^2$, is enough to significantly heat the excitons, even though the total number of hot carriers compared to the total number of excitons is negligible. This indicates that any experiments which purport to have excitons in similar structures at temperatures well below 1 K must be critically examined, because it may be even harder to reach low temperatures than one might think. In any case, we see that in the present structures there is good thermal equilibrium of the excitons above $T= 5 $~K. 

\begin{figure}
\includegraphics[width=0.4\textwidth]{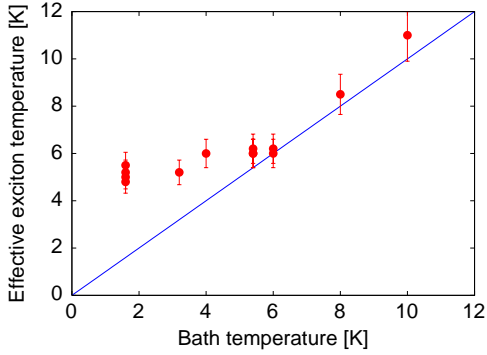}
\caption{The effective temperature of the exciton gas at late times, as deduced from the late-time cloud size. The instrumental spatial resolution has been deconvolved from the data by simple quadrature.}
\end{figure}

In principle, one can reach Bose-Einstein condensation of excitons even at these temperatures, by increasing the exciton density. As the density increases, however, exciton-exciton interaction leads to a strong renormalization of the trapping potential. Fig.~5 shows the spectral position and spectral width of the indirect exciton luminesence at the center of the trap as a function of time. As seen in the lower plot, there is a blue shift of the PL spectrum which is linear with total exciton density. This is consistent with theoretical predictions which take into account first-order and second-order effects such as Hartree-Fock, exchange, and screening, and in principle allows a calibration of the exciton density by comparing the blue shift and spectral width \cite{zimm}.

We conclude that we have truly equilibrium thermodynamic conditions for the excitons in the trap, and above $T= 5$~K, the exciton temperature reaches the lattice temperature, which is the same as the helium bath temperature. This indicates that in-plane traps are a promising approach for Bose-Einstein condensation of excitons. Future work must take into account the strong renormalization of the trapping potential due to the dipole-dipole repulsion of the indirect excitons.

\begin{figure}
\begin{center}
\includegraphics[width=0.45\textwidth]{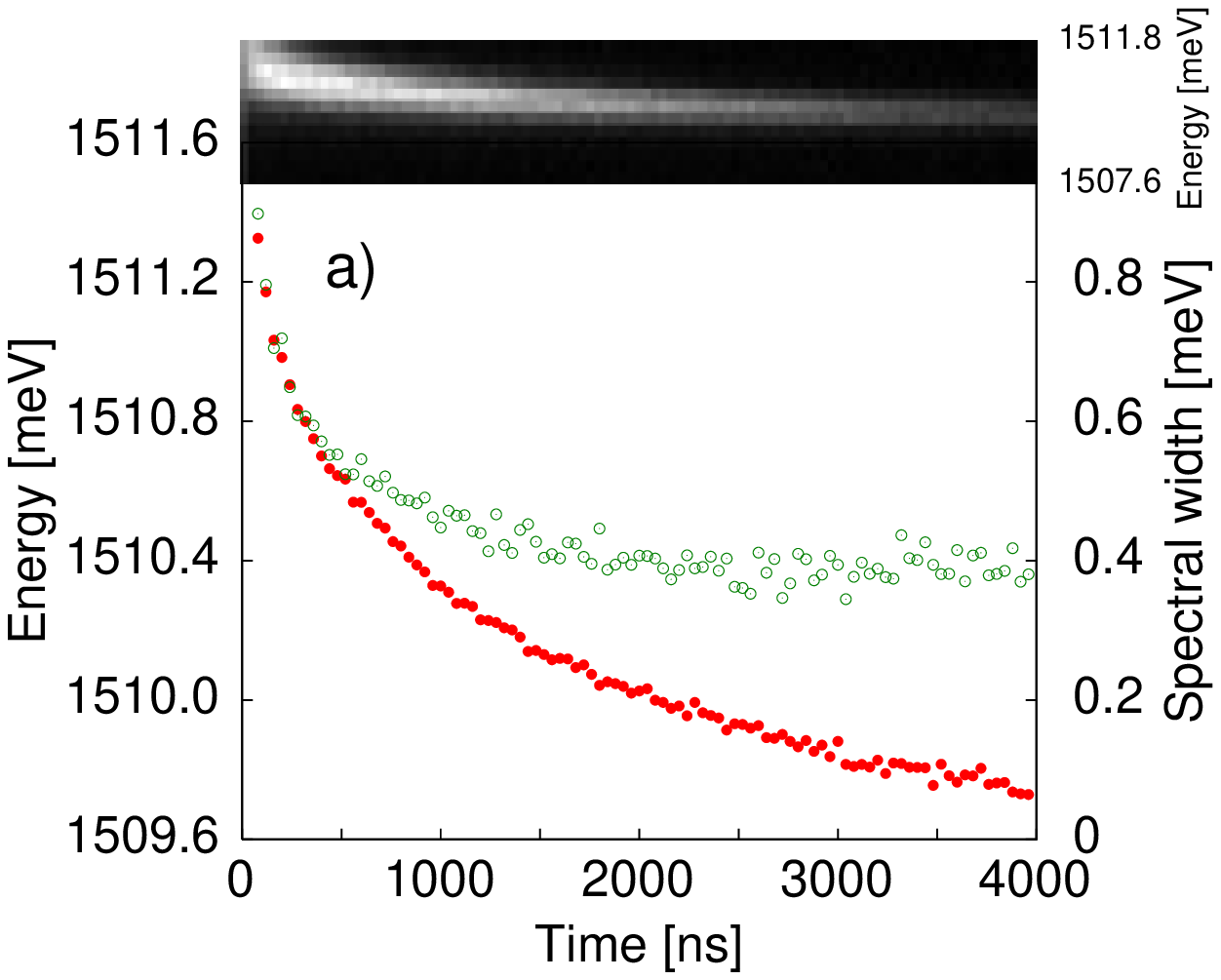}
\includegraphics[width=0.45\textwidth]{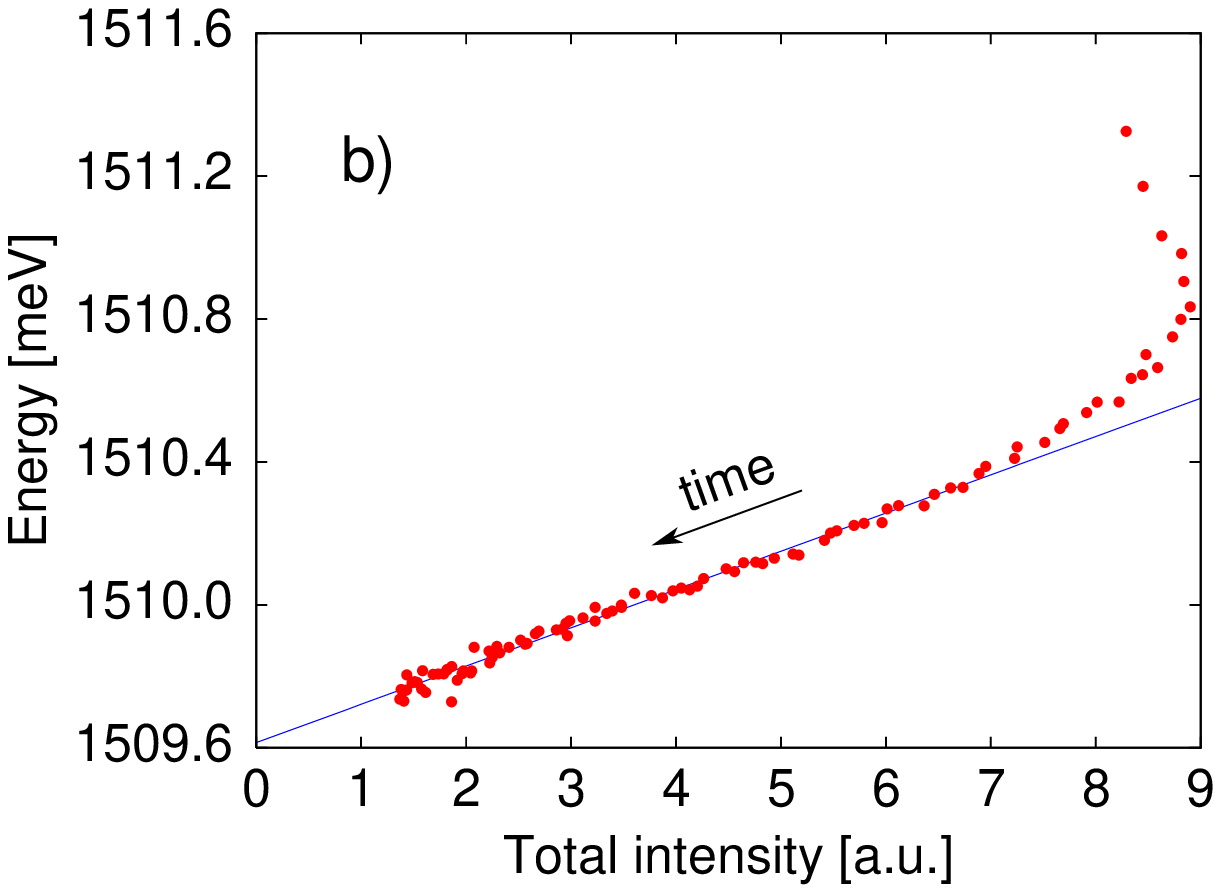}
\end{center}
\caption{a) top image: the luminescence spectrum as a function of time. Graph: the spectral position (solid circles) and spectral width (open circles) as a function of time. b) the spectral position as a function of the total spectrally-integrated PL intensity at $r=0$ at each time, for the same set of data. The spectrally integrated PL intensity is proportional to the exciton density when the lifetime is constant, as shown in Fig. 3. At early times the spectral width and the lifetime are rapidly changing and the photoluminescence intensity is not proportional to the exciton density.}
\end{figure}

{\bf Acknowledgements}. This work has been supported by the National Science Foundation under
grant DMR-0404912 and by the Department of Energy under grant DE-GF02-99ER45780. We thank Vincent Hartwell for calculations of the exciton-phonon thermalization.


\begin{thebibliography}{99}

\bibitem{science-review} For a general review of theory and experiments on Bose condensation of excitons, see D. Snoke, Science {\bf 298}, 1368 (2002).

\bibitem{coh} J. Fernandez-Rossier, C. Tejedor, and R. Merlin, Solid State Comm. 108, 473 (1998).

\bibitem{levitov} J. Keeling, L.S. Levitov, and P.B. Littlewood, Phys. Rev. Lett. {\bf 92}, 176402 (2004).
 
\bibitem{tejedor} J. Fernandez-Rossier, C. Tejedor, Phys. Rev. Lett. 78, 4809 (1997).

\bibitem{yudson} Yu. E. Lozovik and V. I. Yudson, JETP Lett. 22, 274 (1975); Sov. Phys. JETP 44, 389 (1976).
 
\bibitem{littlewood-mag} A.V. Balatsky, Y.N. Joglekar, and P.B. Littlewood, Phys. Rev. Lett.  {\bf 93}, 266801 (2004).

\bibitem{indirect} A. Alexandrou, J. A. Kash, E. E. Mendez, M. Zachau, J. M. Hong, T.
Fukuzawa, and Y. Hase, Phys. Rev. B \textbf{42}, R9225 (1990).

\bibitem{szymanska} M.H. Szymanska and P.B. Littlewood, Phys. Rev. B \textbf{67} 193305
(2003).

\bibitem{diffprl} Z. V\"or\"os, R. Balili, D.W. Snoke, L. Pfeiffer, and K. West, Phys. Rev. Lett. {\bf 94}, 226401
(2005).

\bibitem{gart}  A. G\"artner,   D. Schuh, and J.P. Kotthaus, ``Dynamics of Long-Living Excitons in Tunable Potential Landscapes,'' http://arxiv.org/pdf/cond-mat/0509142.

\bibitem{apl} V. Negoita, D.W. Snoke, and K. Eberl,  Applied Physics
Letters {\bf 75}, 2059 (1999).


\bibitem{eisenstein} M. Kellogg, J.P. Eisenstein, L.N. Pfeiffer, and K.W. West,  Phys. Rev. Lett. {\bf 93}, 036801
(2004)

\bibitem{ibm} E. Tutuc, M. Shayegan, and D.A. Huse, Phys. Rev. Lett. {\bf 93}, 036802 (2004).

\bibitem{bectheory} V. Bagnato and D. Kleppner, Phys. Rev. A {\bf 44}, 7439 (1991).

\bibitem{Trate} J. Feldmann, G. Peter, E. O. G\"obel, P. Dawson, K. Moore, C. Foxon, and R. J. Elliott, Phys. Rev.
Lett. {\bf 59}, 2337 (1987).

\bibitem{ivanov} A.L. Ivanov, J. Phys.: Cond. Mat. {\bf 16}, S3629 (2004). 

\bibitem{snokebc} D.W. Snoke, D. Braun, and M. Cardona, Phys. Rev. B {\bf 44}, 2991 (1991).

\bibitem{zimm} R. Zimmermann, Proc. Int. Conf. Nonlinear Optics and Excitation Kinetics in Semiconductors, in press.

 

\end{thebibliography}
\end{document}